\documentclass[twocolumn,american,prl]{revtex4}
\usepackage{graphicx}
\usepackage{amssymb}


\begin{document}

\title{Discrete-symmetry vortices as angular Bloch modes}

\author{Albert Ferrando$^{1,2}$}

\affiliation{$^{1}$Departament de Matem\`{a}tica Aplicada, Universitat Polit\`{e}cnica
de Val\`{e}ncia. Cam\'{\i} de Vera, s/n. E-46022 Val\`{e}ncia,
Spain.}

\affiliation{$^{2}$Departament d'\`{O}ptica, Universitat de Val\`{e}ncia. Dr.
Moliner, 50. E-46100 Burjassot (Val\`{e}ncia), Spain.}

\date{\today}

\begin{abstract}
The most general form for symmetric modes of nonlinear discrete-symmetry
systems with nonlinearity depending on the modulus of the field is
presented. Vortex solutions are demonstrated to behave as Bloch modes
characterized by an angular Bloch momentum associated to a periodic
variable, periodicity being fixed by the order of discrete point-symmetry
of the system. The concept of angular Bloch momentum is thus introduced
to generalize the usual definition of angular momentum to cases where
$O(2)$-symmetry no longer holds. The conservation of angular Bloch
momentum during propagation is demonstrated.
\end{abstract}


\maketitle

Rotational symmetry in the $x-y$ plane, defined by the $O(2)$ symmetry
group, implies the conservation of the $z$-component of angular momentum
in two-dimensional nonlinear systems. If the $O(2)$ continuous symmetry
is substituted by a discrete rotational symmetry, the usual approach
based on Noether's theorem fails since infinitesimal symmetry transformations
are no longer allowed. Conservation of angular momentum is expected
to breakdown in its usual form and, consequently, nonlinear dynamics
should show special features as compared to the $O(2)$-symmetric
case. Since angular momentum is expected to realize differently in
discrete-symmetry systems, it is particularly interesting to compare
the different behavior of $O(2)$-symmetric and discrete-symmetry
vortices. A general feature of vortex solutions is their characteristic
phase dislocation, which is determined by an integer number that will
be referred to as vorticity (also known as winding-number, {}``topological
charge'' or spin). In $O(2)$-symmetric systems, rotationally invariant
vortices (i.e., vortex solutions whose amplitude is $O(2)$-symmetric)
are eigenfunctions of the angular momentum operator with eigenvalue
(angular momentum) given by vorticity. Oppositely, discrete-symmetry
vortices have no well-defined angular momentum. Examples of both types
of vortices can be found in optics as well as in Bose-Einstein condensate
(BEC) systems. Optical vortices with $O(2)$-symmetric have been experimentally
observed in homogeneous self-defocussing nonlinear Kerr media \cite{swartzlander-prl69_2503}
whereas discrete-symmetry optical vortices have been observed in optically-induced
photonic lattices \cite{neshev-prl92_123903,fleischer-prl92_123904}.
The influence of discrete symmetry in the features of optical vortices
are strongly reflected in their angular properties. The phase of a
discrete-symmetry vortex presents, besides the typical linear angular
dependence characteristic of $O(2)$-symmetric vortices, an additional
sinusoidal contribution completely fixed by the order of the discrete-symmetry
of the system (the so-called group phase)\cite{ferrando-oe12_817}.
Besides, these systems exhibit a vorticity cutoff equally determined
by the order of the discrete-symmetry, in such a way that no vortices
of arbitrary order are permitted \cite{ferrando-arXiv:nlin_0411005}. 

The aim of this paper is two-fold. On the one hand, I will propose
a new approach to the search for vortex solutions of a general nonlinear
equation that describes not only the conventional optical and BEC
cases but also more general situations. I will prove a general expression
for symmetric vortex solutions by showing that they are angular Bloch
modes. On the second hand, I will extend the conservation of angular
momentum to the discrete-symmetry case by introducing the concept
of angular Bloch momentum and by demonstrating its conservation during
propagation.

\begin{figure}
\includegraphics[%
  scale=0.3]{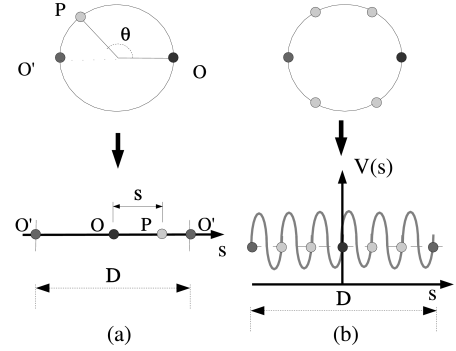}

\caption{(a) Mapping of the angular variable $\theta$ into the non-compact
coordinate $s$. (b) Equivalent periodic potential using this mapping
(for an operator of the type $\tilde{L}(s)=\nabla^{2}-V(s)$) corresponding
to a system with point-symmetry of order $n=6$\label{cap:mapping}.}
\end{figure}

Let us consider the 2D nonlinear eigenvalue equation for a system
possessing discrete point-symmetry of the $\mathcal{C}_{n}$ type
(discrete rotations of order $n$):\begin{equation}
\left[L_{0}+L_{\mathrm{NL}}(|\phi|)\right]\phi=-\mathcal{E}\phi,\label{eq:eigenvalue_equation}\end{equation}
where $L_{0}$ stands for the field-independent part of the full differential
operator (linear part) and $L_{\mathrm{NL}}(|\phi|)$ for the nonlinear
part. There is no restriction about the specific form of $L_{0}$
and $L_{\mathrm{NL}}$. Usually, $L_{0}$ depends on the transverse
position coordinates through gradient operators and explicit functions
of position (a typical example is $L_{0}=\nabla^{2}-V_{0}(x,y)$).
The only assumption made is that the nonlinear part $L_{\mathrm{NL}}$
depends on the field through its modulus exclusively. There can be
also a explicit dependence on coordinates through other functions
(e.g., inhomogeneous nonlinear coefficients: $L_{\mathrm{NL}}=\gamma_{2}(x,y)|\phi|^{2}+\gamma_{4}(x,y)|\phi|^{4}$).
It is considered that the system is invariant under discrete rotations
$\mathcal{C}_{n}$, so that all operators and coefficients defining
$L_{0}$ and $L_{\mathrm{NL}}$ are assumed to be invariant under
discrete rotations of $n$th order. If one is interested in symmetric
solutions, i.e., those functions verifying that their modulus is a
$\mathcal{C}_{n}$-invariant ($|\phi(r,\theta+2\pi/n)|=|\phi(r,\theta)|$),
then the entire operator $L=L_{0}+L_{\mathrm{NL}}$ becomes invariant
under discrete rotations: $L(r,\theta)=L(r,\theta+\epsilon$), $\epsilon\equiv2\pi/n$.
One can transform this angular dependence into a dependence on a length
variable ---defined within an interval of the real axis--- by using
the mapping $\theta\rightarrow s=\theta D/2\pi$. This mapping maps
the unit circle $\mathcal{S}_{1}$ (where $\theta$ is defined: $\theta\in\left[-\pi,\pi\right]$)
onto the real axis interval $[-D/2,D,/2]$ (see Fig.\ref{cap:mapping}).
In terms of the new length variable $s$, invariance under discrete
rotations of the $L$-operator becomes a periodicity property: if
$\tilde{L}(r,s)\equiv L(r,2\pi s/D)$ then $\tilde{L}(r,s+a)=\tilde{L}(r,s)$
where $a$ is the period in the non-compact variable $s$. According
to the previous mapping, the period $a$ is fixed by the discrete-rotation
angle $\epsilon$ and thus it depends on the order of the discrete
symmetry: $a=\epsilon D/2\pi=D/n$. In this way, Eq.(\ref{eq:eigenvalue_equation})
for symmetric functions is transformed into $\tilde{L}(r,s)\varphi(r,s)=-\mathcal{E}\varphi(r,s)$
where $\varphi(r,s)\equiv\phi(r,2\pi s/D)$. One has thus converted
the original discrete-symmetry eigenvalue problem into a problem of
finding the self-consistent modes of a periodic operator. The modulus
of these modes are invariant under periodic translations ($|\varphi(r,s+a)|=|\varphi(r,s)|$),
a consequence of the invariance of the modulus of a symmetric solution
under discrete-rotations. As a clarifying example, one can consider
an standard operator of the type $\tilde{L}(r,s)=\nabla^{2}-V(r,s)$,
where $V(r,s)=V_{0}(r,s)+V_{\mathrm{NL}}(|\varphi(r,s)|)$. If $\varphi(r,s)$
is a solution of $\tilde{L}\varphi=-\mathcal{E}\varphi$ with periodic
modulus then it has to be en eigenmode of the periodic operator generated
by itself $\tilde{L}(r,s)=\nabla^{2}-V(r,s)$. According to Bloch's
theorem, since $V$ is periodic $V(r,s+a)=V(r,s)$ the solution of
an eigenvalue equation of the previous type is given by Bloch modes
in the periodic variable $s$. This argument equally applies to the
general periodic operator $\tilde{L}(r,s)$, so that the most general
solution of $\tilde{L}\varphi=-\mathcal{E}\varphi$ is given by 1D
Bloch modes \cite{ashcroft76}: $\varphi_{p\nu}(r,s)=e^{ips}u_{p\nu}(r,s)$,
where $p$ is the Bloch momentum or pseudo-momentum of the mode and
$u_{p}$ is the so-called Bloch function. The index $\nu$ is known
as the band index and occurs because, at a given value of $p$, there
are many different eigenmodes of $\tilde{L}$. Since $\nu$ does not
play a role in the current discussion, I will omit it from the notation
(although one has to recall that it is always present). The Bloch
function is a periodic function of $s$ with the periodicity of the
$\tilde{L}$ operator : $u_{p}(r,s+a)=u_{p}(r,s)$.

One can obtain interesting properties of solutions of Eq.(\ref{eq:eigenvalue_equation})
by re-interpreting well-known properties of 1D Bloch modes. A Bloch
mode is characterized by the value of its Bloch momentum $p$. However,
unlike plane waves, $p$ is not an eigenvalue of the momentum operator
and its value is constrained to lie in a restricted interval, called
the Brillouin zone, defined by the condition $|p|\le\pi/a$. Moreover,
due to the definition of the length coordinate $s$ in terms of the
angular variable $\theta$, the $\varphi_{p}$ function has to be
additionally periodic in the interval length $D$ ($s\in[-D/2,D/2]$),
$\varphi_{p}(r,s)=\varphi_{p}(r,s+D)$. As a consequence, a discretization
condition for the Bloch momentum $p$ is obtained: $e^{ipD}=1\Rightarrow p=p_{m}=2\pi m/D$,
$m\in\mathbb{Z}$. If one reverses the $\theta\rightarrow s$ mapping
by re-introducing the angular variable $\theta$ in $\varphi_{p}(r,s)=e^{ips}u_{p}(r,s)$,
one obtains the solution in its original form:\begin{equation}
\phi_{m}(r,\theta)=e^{im\theta}\tilde{u}_{m}(r,\theta),\,\,\,\, m\in\mathbb{Z},\,\,\,\,|m|\le n/2.\label{eq:bloch_mode_angular}\end{equation}
I shall call these solutions angular Bloch modes. They constitute
the general symmetric solutions of the nonlinear eigenvalue equation
(\ref{eq:eigenvalue_equation}). Because of the angular nature of
the $\theta$ variable and its relation to the Bloch momentum $p_{m}=2\pi m/D$,
the index $m$ will be referred to as the angular Bloch momentum (or
pseudo angular-momentum) of the angular Bloch mode. The restriction
on the permitted values of the angular Bloch momentum $m$ is a consequence
of the Brillouin zone limitation. The existence of the Brillouin zone
for the Bloch momentum $p$ establishes a condition for its permitted
values: $|p|\le\pi/a$. Since $p$ is discretized according to $p_{m}=2\pi m/D$,
$m\in\mathbb{Z}$, the Brillouin zone limitation effectively imposes
a strict constraint into the angular Bloch momentum $|m|\le D/2a$.
Inasmuch as the period $a$ is determined by the order of symmetry
$n$ ($a=D/n$), one finds that the modulus of the angular Bloch momentum
presents the upper bound $|m|\le n/2$ occurring in Eq.(\ref{eq:bloch_mode_angular}). 

\begin{figure}
\includegraphics[%
  scale=0.4]{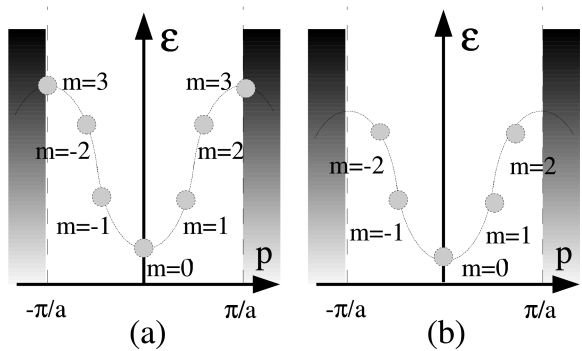}

\caption{Schematic representation of Brillouin zones for two different order
of symmetries: (a) even $n$ ($n=6$); (b) odd $n$ ($n=5$). Permitted
angular Bloch modes are shown.\label{cap:Brillouin_zones}}
\end{figure}

The expression (\ref{eq:bloch_mode_angular}) is the same one that
it is obtained for vortex solutions in a nonlinear system with point-symmetry
$\mathcal{C}_{n}$ using group theory arguments \cite{ferrando-oe12_817,ferrando-arXiv:nlin_0411005}:
$\phi_{\bar{l}}(r,\theta)=e^{i\bar{l}\theta}\phi_{0}^{(\bar{l})}(r,\theta)$,
$\bar{l}\in\mathbb{Z}$ being the index of the group representation
where the solution belongs to and $\phi_{0}^{(\bar{l})}$ a discrete-rotation
invariant function, $\phi_{0}^{(\bar{l})}(r,\theta+2\pi/n)=\phi_{0}^{(\bar{l})}(r,\theta)$.
The invariant function $\phi_{0}^{(\bar{l})}$ predicted by group
theory is nothing but the Bloch function $\tilde{u}_{m}$ of the angular
Bloch mode (\ref{eq:bloch_mode_angular}). Vorticity equals the index
of the representation, so that $\phi_{\bar{l}}$ represents a vortex
with vorticity $v=\bar{l}$ (the exception is the $\bar{l}=n/2$ ---even
$n$--- solution)\cite{ferrando-arXiv:nlin_0411005}. The comparison
of $\phi_{\bar{l}}(r,\theta)=e^{i\bar{l}\theta}\phi_{0}^{(\bar{l})}(r,\theta)$
and Eq.(\ref{eq:bloch_mode_angular}) yields to interesting equivalences.
Vortex solitons of vorticity $v=\bar{l}$ appear now as angular Bloch
modes carrying angular Bloch momentum $m=v=\bar{l}$ ($m=n/2$ ---even
$n$--- excluded). The restriction on angular Bloch momentum $|m|\le n/2$,
along with the previously established relation between vorticity and
angular Bloch momentum, fixes a cutoff for the vorticity: $|v|\le n/2$
($v=n/2$ ---even $n$--- excluded). This condition can be also rephrased
as:\begin{equation}
\left|v\right|<n/2\,\,\,(\mathrm{even}\,\, n)\,\,\,\mathrm{and}\,\,\,\left|v\right|\le(n-1)/2\,\,\,(\mathrm{odd}\,\, n).\label{eq:cutoff}\end{equation}
The explicit distinction between even and odd orders is made to remark
the different behavior existent in the corresponding Brillouin zones
in both cases. For odd $n$, the discretized Bloch momentum $p_{m}$
can never achieved the zone limit $|p_{m}|<\pi/a$, whereas this is
certainly possible for even $n$. This can be clearly visualized in
Fig. \ref{cap:Brillouin_zones}, where two examples of Brillouin zones
for even and odd $n$ are shown. It is remarkable that the vorticity
cutoff expressed in Eq.(\ref{eq:cutoff}) has been obtained in a completely
different mathematical framework using group-theory arguments \cite{ferrando-arXiv:nlin_0411005}. 

At this point, it is interesting to compare the discrete-symmetry
situation with that corresponding to continuous rotational symmetry.
In the $O(2)$ case, Noether's theorem ensures the conservation of
the $z$- component of the angular momentum associated to the $\phi$
field during propagation:\begin{equation}
j_{z}=\int_{\mathbb{R}^{2}}dxdy\,\phi^{*}(\mathbf{r})(\mathbf{r}\wedge\nabla)_{z}\phi(\mathbf{r})\Rightarrow\frac{dj_{z}(z)}{dz}=0.\label{eq:angular_momentum}\end{equation}
This quantity is also the expectation value of the angular momentum
operator $L_{z}=-i\partial/\partial\theta$ for the $\phi$ field,
i.e., $j_{z}=\left\langle \phi|L_{z}|\phi\right\rangle $. Since $O(2)$-symmetric
vortex solutions are of the type $\phi_{\bar{l}}(r,\theta)=e^{i\bar{l}\theta}f_{\bar{l}}(r)$,
vortex solutions are eigenfunctions of the angular momentum operator
with eigenvalue $\bar{l}$. Therefore, $j_{z}$ equals $\bar{l}$
for (normalized to one) vortex solutions and both represent the same
constant of motion. On the other hand, the vorticity $v$ of $O(2)$-symmetric
vortices is directly given by $\bar{l}$: $v=j_{z}=\bar{l}$. Thus,
angular-momentum and vorticity are equivalent quantities for $O(2)$-symmetric
vortices.

In a discrete-symmetry system the $z$-component of angular momentum
$j_{z}$ (Eq.(\ref{eq:angular_momentum})) is no longer conserved.
Noether's theorem does not apply because the symmetry under consideration
ceases from being continuous. It is then necessary to find a different
conserved quantity that explicitly manifests the discrete-symmetry
invariance of the system. It will be proved next that this quantity
is nothing but the angular Bloch momentum. In order to carry out the
previous demonstration, it is simpler to resort to the representation
of Bloch modes in terms of the non-compact variable $s$. It will
be assumed that evolution is provided by a first-order operator in
the evolution variable $z$, i.e., $\tilde{L}\varphi(r,s,z)=-i\partial\varphi(r,s,z)/\partial z$.
This type of evolution operator includes Schroedinger-like equations
(like those appearing in nonlinear paraxial evolution in optical systems
or in the evolution of Bose-Einstein condensates) but also more general
evolution equations, as non-paraxial forward equations in axially-invariant
optical systems \cite{ferrando-arXiv:physics/0407029}. My aim now
is to analyze the evolution of a field amplitude whose initial condition
is given by a Bloch mode with well-defined pseudo-momentum $p$, i.e.,
$\varphi(r,s,0)=\varphi_{p}(r,s)=e^{ips}u_{p}(r,s)$. It is important
to stress that this mode does not need to be a stationary solution
of the vortex type satisfying the nonlinear eigenvalue Eq.(\ref{eq:eigenvalue_equation}),
so that its amplitude can evolve with $z$. It is a known fact that
Bloch momentum is conserved during propagation in a periodic linear
system. However, this question is not obvious in the nonlinear case
inasmuch as the evolution operator explicitly depends on the field
itself.

Let us consider the total evolution from 0 to $z$ as a succession
of infinitesimal evolution steps of length $\varepsilon$ in the limit
$\varepsilon\rightarrow0$. In a first-order evolution equation, the
infinitesimal evolution of a field amplitude from the axial slice
$z_{j}$ to the following slice $z_{j+1}=z_{j}+\varepsilon$ is given
by: \begin{equation}
\varphi_{j+1}(r,s)=e^{i\tilde{L}(|\varphi_{j}|)\varepsilon}\varphi_{j}(r,s).\label{eq:evolution}\end{equation}
Now I will prove that if $\varphi_{0}$ ($\varphi_{0}\equiv\varphi(r,s,0)$)
is a Bloch mode with well-defined Bloch momentum $p$, then $\varphi_{j}$
is also a Bloch mode with the same pseudomentum $p$ for all values
of $j$. The proof is carried out by induction. Let us start by calculating
the pseudo-momentum of the $j=1$ amplitude $\varphi_{1}$. The initial
amplitude $\varphi_{0}$ is a Bloch mode with pseudo-momentum $p$
($\varphi_{0}=\varphi_{p}(r,s)=e^{ips}u_{p}(r,s)$) and thus it is
an eigenfunction of the translation operator $T_{a}$ : $T_{a}\varphi_{0}(r,s)=\varphi_{0}(r,s+a)=e^{ipa}\varphi_{0}(r,s)$.
The amplitude $|\varphi_{0}|$ is a periodic function of $s$ with
periodicity $a$ since it is the modulus of the periodic Bloch function
$u_{p}(r,s)$: $|\varphi_{0}(r,s+a)|=|\varphi_{0}(r,s)|$, so that
the nonlineal part of the $\tilde{L}$ operator is translational invariant,
$L_{\mathrm{NL}}(|\varphi_{0}(r,s+a)|)=L_{\mathrm{NL}}(|\varphi_{0}(r,s)|)$.
Therefore the full operator $\tilde{L}$ is invariant under finite
translations since both its linear and nonlinear part are invariant,
$[\tilde{L}(|\varphi_{0}|),T_{a}]=0$. This implies in turn that the
infinitesimal evolution operator in Eq.(\ref{eq:evolution}) commutes
with $T_{a}$: $[e^{i\tilde{L}(|\varphi_{0}|)\varepsilon},T_{a}]\stackrel{\varepsilon\rightarrow0}{=}[1+i\tilde{L}(|\varphi_{0}|)\varepsilon,T_{a}]=0$.
Now I apply the $T_{a}$ operator onto $\varphi_{1}$ and take into
account Eq.(\ref{eq:evolution}) for $j=0$ to find $T_{a}\varphi_{1}=e^{i\tilde{L}(|\varphi_{0}|)\varepsilon}T_{1}\varphi_{0}=e^{ipa}\varphi_{1}$,
where I have used the fact that the evolution and translation operator
commute between them and that $\varphi_{0}$ is an eigenfunction of
$T_{a}$. Consequently, $\varphi_{1}$ is a Bloch mode with pseudo-momentum
$p$. Now, following the induction procedure, I will assume the Bloch-mode
property to be true for $\varphi_{j}$ and I will prove it for $\varphi_{j+1}$.
The demonstration is analogous to the $j=0$ case. I assume $\varphi_{j}$
is a Bloch-mode with pseudo-momentum $p$. Then its modulus is translational
invariant, so $\tilde{L}(|\varphi_{j}|)$ and $e^{i\tilde{L}(|\varphi_{j}|)\varepsilon}$
are: $[\tilde{L}(|\varphi_{j}|),T_{a}]=[e^{i\tilde{L}(|\varphi_{j}|)\varepsilon},T_{a}]=0$.
If one acts with the translation operator $T_{a}$ onto $\varphi_{j+1}$,
commutes the translation and evolution operators and takes into account
that $T_{a}\varphi_{j}=e^{ipa}\varphi_{j}$ (Bloch-mode condition
for $\varphi_{j}$), one readily finds that $T_{a}\varphi_{j+1}=e^{ipa}\varphi_{j+1}$.
This shows that $\varphi_{j}$ is a Bloch-mode with pseudo-momentum
$p$ for all values of $j$ if the initial amplitude $\varphi_{0}$
is. Expressed in different words: axial nonlinear evolution preserves
the Bloch momentum $p$.

The continuous limit $\varepsilon\rightarrow0$ of the previous statement
permits to give an expression for the evolving field amplitude $\varphi(r,s,z)$:
if $\varphi(r,s,0)=e^{ips}u_{p}(r,s)$ then it is also true that $\varphi(r,s,z)=e^{ips}u_{p}(r,s,z)$,
$u_{p}$ being a $z$-dependent Bloch function. If we reintroduce
the angular variable $\theta$ instead of the non-compact one $s$,
one obtains that the general expression for the evolution of an initial
angular Bloch mode will well-defined angular Bloch momentum $m$ in
the system under consideration is:\begin{equation}
\phi_{m}(r,\theta,z)=e^{im\theta}\tilde{u}_{m}(r,\theta,z),\,\,\,\, m\in\mathbb{Z},\,\,\,\,|m|\le n/2.\label{eq:angular_Bloch_mode_evolution}\end{equation}
This equation establishes that the angular Bloch momentum $m$ is
preserved by nonlinear evolution.

The interplay between ordinary angular momentum and angular Bloch
momentum can be visualized by calculating the expectation value of
the angular momentum operator for the Bloch mode amplitude (\ref{eq:angular_Bloch_mode_evolution}).
One easily finds that $j_{z}(z)=m+\left\langle \tilde{u}_{m}|L_{z}|\tilde{u}_{m}\right\rangle (z)$
or\begin{equation}
m=j_{z}(z)-\left\langle \tilde{u}_{m}|L_{z}|\tilde{u}_{m}\right\rangle (z).\label{eq:Bloch_angular_momentum_conservation}\end{equation}
This property shows that angular Bloch momentum is conserved despite
angular momentum is not. The expectation value appearing in Eq.(\ref{eq:Bloch_angular_momentum_conservation})
is $j_{u}^{m}(z)\equiv\int_{\mathbb{R}^{2}}\tilde{u}_{m}^{*}(-i\partial/\partial\theta)\tilde{u}_{m}$
corresponding to the angular momentum associated to the $\theta$-dependent
Bloch functions $\tilde{u}_{m}$. The conservation of the angular
Bloch momentum is then the result of a subtle balance between two
non-conserved quantities: the conventional angular momentum $j_{z}(z)$
and the angular momentum $j_{u}^{m}(z)$ related to Bloch functions.
The latter can be attributed to the presence of the discrete-symmetry
system that acts as an angular periodic crystal. The dependence of
the $\tilde{u}_{m}$ Bloch functions on $\theta$ is a consequence
of the angular periodicity of the system. In order to see it, it is
clarifying to consider an $O(2)$-invariant medium as the $n\rightarrow\infty$
limit of a system with discrete symmetry of $n$th order. In such
a case, the $\tilde{u}_{m}$ Bloch functions become independent of
the $\theta$ angle and, therefore, their associated angular momentum
vanishes $j_{u}^{m}(z)\stackrel{n\rightarrow\infty}{\rightarrow}0$.
As a consequence, angular momentum becomes angular Bloch momentum
$j_{z}(z)\stackrel{n\rightarrow\infty}{\rightarrow}m$ and, therefore,
a constant of motion. It is then reasonable to attribute the angular
momentum contribution $j_{u}^{m}(z)$ to the discrete-symmetry system
acting as an angular periodic crystal. One can interpret Eq.(\ref{eq:Bloch_angular_momentum_conservation})
as the statement that during propagation conventional angular momentum
is transfered to the discrete-symmetry system (the angular crystal),
and vice-versa, in such a way angular Bloch momentum is conserved:
$dj_{z}(z)/dz=-dj_{u}^{m}(z)/dz$. 

All the previous arguments apply to every type of evolving field provided
it satisfies the initial condition of having well-defined angular
Bloch momentum. Conservation of angular Bloch momentum appears then
as a property which occurs independently whether the solution is stationary
or not. Nonlinear stationary solutions of Eq.(\ref{eq:eigenvalue_equation})
are only particular cases to which the general conservation law can
be applied to. For discrete-symmetry vortices the angular-momentum-vorticity
equivalence $v=j_{z}$ of $O(2)$-invariant solutions is lost. Since
discrete-symmetry vortices are angular Bloch modes of vorticity $v=m$,
Eq.(\ref{eq:Bloch_angular_momentum_conservation}) establishes a new
relation between vorticity and the angular momentum carried by the
vortex field (as defined in Eq.(\ref{eq:angular_momentum})): $v=j_{z}-j_{u}^{m}$.
In this way, the angular Bloch momentum concept permits to unveil
the role played by angular momentum and vorticity in systems whose
symmetry is no longer $O(2)$ but a point-symmetry of lesser order.

The author is thankful to P. F. de C\'{o}rdoba and M. Zacar\'{e}s
for useful discussions. This work was financially supported by the
Plan Nacional I+D+I (grant TIC2002-04527-C02-02), MCyT (Spain) and
FEDER funds. Author also acknowledges the financial support from the
Generalitat Valenciana (grants GV04B-390 and Grupos03/227).

\end{document}